# Recording zero-point current and voltage fluctuations

G B Lesovik

In this paper we consider various methods for measuring current fluctuations. Our aim is to reveal a quantity which can be measured in treating fluctuations and transfer statistics on the whole. The answer is known for an average current where the quantity sought for is the averaged current operator $\langle I \rangle = \mathrm{Tr}\{\rho I\}$ in view of the validity of the ergodic hypothesis. The situation with current–current correlators is far less clear, since the operators should be arranged in time (in general, current operators at different times are not commutative).

In fact, this problem is reduced to the measurement of vacuum current fluctuations, and similar to the one of recording photons in optics and the measurement of vacuum electromagnetic fluctuations, although there is a significant difference.

Considerable recent attention has been focussed on the measurement of zero-point current and voltage oscillations including the very possibility of such measurements. Current fluctuations have been studied both theoretically and experimentally [1–5]. In Refs [6, 7] measurements were performed at frequencies at which zero-point oscillations can arise at practically attainable temperatures ($\hbar\omega > k_\mathrm{B}T$).

On the other hand, paper [8] attracted considerable interest to the possible breaking of the phase of conducting electrons by vacuum fluctuations of an electric field, which can significantly modify the localization behaviour at zero temperature.

To begin, let us consider the spectral density of fluctuations.

**1. Measurement of spectral density of noise with a resonance circuit.** Current fluctuations of finite frequency are usually measured by one of two main methods. In the first, the current is measured as a time-dependent function $I(t)$, for example, with a normal ammeter, and then the spectral density $S(\omega)$ is calculated numerically using a Fourier transformation.

The classical equation of motion for an ammeter coincides with the equation for an oscillator with friction and external force $\propto I(t)$

$$\ddot{\phi} = -\Omega^2 \phi - \gamma \dot{\phi} + \lambda I(t) \,. \tag{1}$$

Making the Fourier transformation, we express the angle-angle correlator as:

$$\langle \phi_\omega \phi_{-\omega} \rangle = \frac{\lambda^2 I_\omega I_{-\omega}}{(\Omega^2 - \omega^2)^2 + \omega^2 \gamma^2} \,. \tag{2}$$

To eliminate proper oscillations, it is usually assumed that $\gamma \gg \Omega_1 = (\Omega^2 - \gamma^2/4)^{1/2}$.

The method is appropriate for recording ultra-low frequency noise, for instance, flicker noise, but, for various reasons, it cannot be used at high frequencies. For example, as in the case of voltage measurements with a discrete voltmeter there is a 'dead' time during which the device cannot record changes in current (below we consider the measurement of a time-dependent current–current correlator with an ammeter).

In recording high frequencies it is more suitable to use a resonance circuit (RC) as a detector coupled by inductance with the investigated conductor so that the RC is not affected by dc.

In this case the detector can still be described by Eqn (1), but now the external force is proportional to the derivative of the measured current $\lambda \dot{I}(t)$, and the circuit quality should be high, so $\gamma \ll \Omega$.

Then the detector response is a changed charge at the capacitor, $\phi \to Q$,

$$Q^2 = \int \mathrm{d}\omega \, \frac{\lambda^2 \omega^2 I_\omega I_\omega}{(\Omega^2 - \omega^2)^2 + \omega^2 \gamma^2} \,. \tag{3}$$

We have considered the same system in quantum-mechanical terms [1], assuming the circuit to have a certain temperature $T_\mathrm{LC}$. Treating the RC as an oscillator with infinitely small damping $\eta$, we have found the correction to squared charge fluctuations, which is of second order with respect to the inductance coupling constant. The result can be formulated as follows: the measurable response of the considered detector at the resonance frequency $\Omega$ is expressed via current correlators as:

$$S_\mathrm{meas} = K\Big\{ S_+(\Omega) + N_\Omega \big[S_+(\Omega) - S_-(\Omega)\big] \Big\} \,, \tag{4}$$

where we introduce the notations

$$S_+(\Omega) = \int \mathrm{d}t \, \langle I(0) I(t) \rangle \exp(\mathrm{i}\Omega t) \,,$$

$$S_-(\Omega) = \int \mathrm{d}t \, \langle I(t) I(0) \rangle \exp(\mathrm{i}\Omega t) \,.$$

The frequency $\Omega$ is assumed to be positive in the expressions, $N_\Omega$ are the Bose occupation numbers of the oscillator, i.e.,

$$N_\Omega = \left[\exp\left(\frac{\hbar\Omega}{k_B T_{LC}}\right) - 1\right]^{-1},$$

$K$ is an effective constant of coupling between the quantum wire and the RC, $\langle A \rangle = \text{Tr}(\rho A)$, where $\rho$ is the electron density matrix, and the time-dependent current operators are determined in the ordinary way as $I(t) = \exp(iHt) I \exp(-iHt)$. The derived expression should be compared with the widely used formula [9]

$$S(\omega) = \int dt \, \exp(i\omega t) \left\langle \frac{1}{2} \{I(0)I(t) + I(t)I(0)\} \right\rangle. \quad (5)$$

Note that the formula includes the symmetrized current correlator. The symmetrization results from the fact that the current operators at different times are not commutative, the symmetrization guarantees the correlators to be Hermitian, and is likely similar to the corresponding classical expression [9]. It is easy to check that the introduced quantities $S_+$ and $S_-$ are also Hermitian [under the condition of time homogeneity $\langle I(t_1)I(t_2) \rangle = f(t_1 - t_2)$, as was assumed in the derivation of Eqn (4)].

Formula (5) leads to the well known expression for the spectral density of fluctuations in an equilibrium conductor [10]:

$$S(\Omega) = 2G\hbar\Omega \left[\frac{1}{2} + \frac{1}{\exp(\hbar\Omega/k_B T) - 1}\right]. \quad (6)$$

This means that at zero temperature the fluctuations should be proportional to frequency, which is usually interpreted as an analog of zero (vacuum) oscillations in an electromagnetic field.

However, as is known from optical measurement, normal photodetectors do not record zero oscillations, because the energy required to excite an atom in the detector cannot be extracted from the vacuum (see, e.g., Ref. [11]). At the same time, zero-point oscillations can be observed (although by a more complicated way than for usual fluctuations) in the Lamb shift of levels [12], in the Casimir effect [13], or with the use of the so-called Mandel quantum counter [14], which is initially prepared in an excited state and hence can record zero oscillations.

Analyzing Eqn (4), we will show that the RC can operate as a photodetector, which is not affected by zero-point oscillations, or a quantum counter, but its response is never described by the standard Nyquist expression (6) as would be expected.

When the detected frequency greatly exceeds the temperature of the LC-detector, $N_\Omega$ is exponentially small, and the only non-vanishing term in Eqn (4) is the positive part of the spectral density $S_+(\Omega)$, which describes the energy emission from the conductor to the RC. In this case the LC-detector works as a normal photoreceiver. As an example, we express $S_+(\Omega)$ in a coherent conductor with transmission coefficient $D$ at zero temperature and finite voltage as

$$S_+(\Omega) = \frac{2e^2}{h} D(1-D)(eV - \hbar\Omega), \quad (7)$$

when $\hbar\Omega < eV$, and $S_+(\Omega) = 0$ in the opposite case. Here we ignore the energy dependence of the transmission coefficient.

Expression (7) coincides with the *excess* spectral density calculated by the symmetrized correlator (5).

If the frequency is much less than the temperature of the RC, i.e. $\hbar\Omega \ll k_B T_{LC}$, the Bose numbers $N_\Omega$ can be replaced by $k_B T_{LC}/\hbar\Omega$. The difference $S_+(\Omega) - S_-(\Omega)$ is *negative*, and for a quantum conductor with a transmission coefficient weakly dependent on energy we obtain

$$S_+ - S_- = -2\hbar\Omega G,$$

where $G = (2e^2/h) \sum_n D_n$ is the conductance.

Note that the singular behaviour of the spectral density at $\hbar\Omega = eV$, which was found in Ref. [3] for the symmetrized expression $S_+ + S_-$, does not take place for $S_+ - S_-$, therefore we can conclude that the only singularity which can be measured at zero temperature and finite voltage is determined by frequency cut-off of voltage in $S_+(\Omega)$.

Now we have for $\hbar\Omega \ll k_B T_{LC}$:

$$S_{meas} = K\{S_+(\Omega) - 2G k_B T_{LC}\}.$$

The meaning of the negative term is clear: the LC-detector is 'cooled', emitting energy into the conductor. Thus, in this limit the zero-point oscillations presented by $S_-$ can be observed in some sense, but the final result includes an expression different from the Nyquist formula (6).

If the conductor is in equilibrium (to an accuracy of a weak interaction with the RC), we have at low frequencies: $S_{meas} \propto 2G(T_e - T_{LC})$. This expression is equal to zero, when the electron temperature $T_e$ is equal to the temperature $T_{LC}$ of the LC-detector, as was to be expected for the total equilibrium of the system.

At intermediate frequencies $k_B T_e, eV \ll \hbar\Omega \ll k_B T_{LC}$, the measuring response is negative: $S_{meas} = -2G k_B T_{LC}$.

When the low-frequency limit $\hbar\Omega \ll eV_{bias}, k_B T$ is considered, it makes no difference whether we use $S_+(\Omega)$ or $S_-(\Omega)$ or the Fourier transform of the symmetrized expression (5) to determine the spectral density, since the result will be the same with the accuracy of small corrections of the order of $\hbar\Omega/eV, k_B T$.

We shall make some remarks on the work [7], whose authors have performed a very interesting experiment on the measurement of the spectral density of fluctuations under a finite bias voltage. Interpreting the results [7] they suggested that the measured spectral density coincided with the Fourier transform of the *symmetrized* current correlator. In our opinion this interpretation is not unique. The point is that the derivative of the spectral density of the current with respect to the bias voltage, which was studied in Ref. [7], is the same (at least at zero temperature) for the symmetrized correlator (5) and for the above-considered $S_+$ from (4). To clear up whether the presented theory can describe the experimental data or not, one should additionally check the magnitude of the partial derivative of the signal with respect to the conductance or measure precisely the absolute value of the noise, which is probably more difficult.

We reiterate that, in our opinion, a measuring scheme where a given frequency is selected with a certain resonance circuit, does not enable one to measure directly zero-point oscillations. For a measuring scheme other than the above RC or a normal ammeter (see below), the question of how the measuring quantity is expressed by the current correlator should be analyzed independently in each particular case.

**2. Measurement of fluctuations of transferred charge with a quantum galvanometer.** Now we consider the question of the contribution of vacuum fluctuations to the transferred charge $Q$ recorded with a quantum galvanometer. Since in calculating the characteristic function determined by analogy with the classical expression as

$$\chi(\lambda) = \left\langle \exp\left\{ i\lambda \int_0^t \hat{I}\, dt' \right\} \right\rangle,$$

one should know how to arrange the current operators in time, Levitov et al. (see Ref. [5]) proposed to use a model galvanometer, i.e. a spin 1/2 rotating in the magnetic field induced by the current in a conductor. The classical current turns the spin through some angle, the quantum current produces a superposition of spin states with different rotation angles; the squared amplitude of the states is considered as the probability of the corresponding charge transfer. Then the characteristic function should correspond to the expression:

$$\chi(\lambda) = \left\langle \tilde{T} \exp\left( \frac{i\lambda}{2} \int_0^t \hat{I}\, dt' \right) T \exp\left( \frac{i\lambda}{2} \int_0^t \hat{I}\, dt'' \right) \right\rangle, \quad (8)$$

where $T$ and $\tilde{T}$ are the forward and backward time-ordering operators respectively. It turns out that for this measurement the average squared fluctuations of the transferred charge, for example in equilibrium, can be calculated by integrating the spectral density of fluctuations determined by the Nyquist formula

$$\langle\langle Q^2(t) \rangle\rangle = \int d\omega\, S_{\mathrm{symm}}(\omega) \int_0^t dt_1 \int_0^t dt_2\, \exp[-i(t_1 - t_2)\omega], \quad (9)$$

and, therefore, vacuum fluctuations can be recorded.

The method described for recording the transferred charge can be carried out in practice† using the μSR technique, although the short life-time of the muon ($\sim 10^{-6}$ s) substantially restricts its potential.

To calculate the average squared fluctuations, we should expand expression (8) up to the second order with respect to $\lambda$ and use the well known formula:

$$\langle Q^2 \rangle = \frac{d^2}{d[i\lambda]^2} \chi(\lambda) \bigg|_{\lambda=0}. \quad (10)$$

Expanding the exponents in Eqn (8) and taking into account the time arrangement, we obtain:

$$\chi(\lambda) = \left\langle \left[ \frac{i\lambda}{2} \right]^2 \left\{ \int_0^t \hat{I}(t')\, dt' \int_0^t \hat{I}(t'')\, dt'' + \int_0^t dt' \int_0^{t'} dt'' \{ \hat{I}(t')\hat{I}(t'') + \hat{I}(t'')\hat{I}(t') \} \right\} \right\rangle. \quad (11)$$

It is seen that the sum of integrals in this expression can be transformed into a double integral of the current correlator symmetrized in time within the limits from 0 to $t$. Then using (10), we arrive at expression (9).

The possibility of recording vacuum fluctuations with a quantum galvanometer can be qualitatively explained by the fact that the spin-detector, which in the initial state has a polarization perpendicular to the magnetic field, does not change its energy during further evolution.

A similar phenomenon seems to arise in the problem of the Josephson contact shunted by a normal conductor, where the energy of the system is also degenerate with respect to the phase difference, and vacuum voltage fluctuations break the phase. The resultant phase difference can be calculated via the symmetrized voltage correlator, which is generally given by the fluctuative-dissipative theorem (FDT). At zero temperature the correlator of the phase difference is

$$\langle \exp\{i\Delta\phi(0) - \Delta\phi(t)\} \rangle \propto \exp\left\{ -\ln(\omega_0 t) \frac{8e^2}{Gh} \right\}$$

$$= \frac{1}{(\omega_0 t)^{8e^2/Gh}}.$$

The considered spin rotation (depolarization) in the magnetic field induced by the current will make a substantial contribution to the *phase breaking at zero temperature* in the weak localization problem, if the fluctuations of the electric field are suppressed. The latter condition is partially fulfilled if the resistance of the external circuit is much less than that of the sample, and the sample voltage fluctuations can be neglected. When the length of the phase breaking is much less than the length of the quasi-one-dimensional sample, the suppression of the fluctuations weakly affects the phase breaking for short scales, where the fluctuations of the electric field act effectively.

**3. Measurement of current correlator with an ammeter.** Now we consider the measurement of time-dependent current correlator with an ammeter. We assume (for simplicity) that the ammeter's current circuit position $x$ is measured not continuously but from time to time. We also assume that, on average, such a measurement weakly affects the density matrix of the electron–ammeter system (below we also refer to the ammeter as the oscillator), and between the measurements the system changes independently from the observer.

The measurement of the coordinate $x = x_0$ will be described by the projection of the density matrix

$$\rho \to |x_0\rangle \frac{\langle x_0|\hat{\rho}|x_0\rangle}{\mathrm{Tr}_e \langle x_0|\hat{\rho}|x_0\rangle} \langle x_0|. \quad (12)$$

An arbitrary orthogonal set $\langle x'|x\rangle \sim \delta_{x'x}$ can be used for measuring states $|x\rangle$ (it can be approximately orthogonal, such as coherent states, but must have a certain rule to determine the magnitude of the measured coordinate). For convenience we require the set to be complete, although it is not necessary.

The coordinate correlator $x$ can be presented as:

$$\langle x(t_1)x(t_2) \rangle = \frac{\mathrm{Tr}_{e1} \int dx\, \langle x|\hat{\rho}_1|x\rangle x}{\mathrm{Tr}_e \int dx\, \langle x|\hat{\rho}_1|x\rangle}$$

$$\times \mathrm{Tr}_{e2} \left\{ \int dx'\, x' \langle x'|\hat{S}_{2,1}|x\rangle \frac{\langle x|\hat{\rho}_1|x\rangle}{\mathrm{Tr}_e\langle x|\hat{\rho}_1|x\rangle} \langle x|\hat{S}_{2,1}^\dagger|x'\rangle \right\}. \quad (13)$$

The limit, in which the measured coordinate correlator corresponds to the current correlator at the same time difference, can be obtained if the electron relaxation time $\tau_{\mathrm{el}}$ greatly exceeds the characteristic time of the oscillator $\tau_x$.

---

† This possibility was pointed out by Prof. G Blatter; the author is thankful to Prof. Yu M Belousov for consultations on the μSR method.

At $t \gg \tau_x$ the main contribution to the correlator is given by

$$\langle x(t_1)x(t_2)\rangle = \left(\frac{i\lambda M}{\hbar}\right)^2 \text{Tr}\left\{\hat{\rho}\int_{-\infty}^{0} dt'' \right.$$
$$\times \int_0^t dt' \hat{I}(t'')\hat{I}(t')\hat{x}(t'')\hat{x}(0)\langle\langle\hat{x}(t')\hat{x}(t) - \hat{x}(t)\hat{x}(t')\rangle\rangle$$
$$\left. - \hat{x}(0)\langle\langle\hat{x}(t')\hat{x}(t) - \hat{x}(t)\hat{x}(t')\rangle\rangle\hat{x}(t'')\hat{I}(t')\hat{I}(t'')\right\}. \quad (14)$$

Here we set $t_1 = 0$, $t_2 - t_1 = t$, $\hat{x}$ is the usual operator of the oscillator coordinate. The double angular brackets denote the diagonalization of the operator $\hat{A} \to \int dx |x\rangle\langle x|\hat{A}|x\rangle\langle x|$.

We use the assumption that the correlations are damped over a rather short period and separate the averaged multiplication of four coordinate operators into pair correlators with a small difference in time. As a result, we transform the expression to

$$\langle x(t_1)x(t_2)\rangle = \left(\frac{i\lambda M}{\hbar}\right)^2 \int_{-\infty}^0 dt'' \int_0^t dt'$$
$$\times \left\{\langle \hat{I}(t'')\hat{I}(t')\rangle_S \langle \hat{x}(t'')\hat{x}(0) - \hat{x}(0)\hat{x}(t'')\rangle \right.$$
$$\times \langle\langle\hat{x}(t')\hat{x}(t) - \hat{x}(t)\hat{x}(t')\rangle\rangle$$
$$+ \langle \hat{I}(t'')\hat{I}(t')\rangle_A \langle \hat{x}(t'')\hat{x}(0) + \hat{x}(0)\hat{x}(t'')\rangle$$
$$\left. \times \langle\langle\hat{x}(t')\hat{x}(t) - \hat{x}(t)\hat{x}(t')\rangle\rangle\right\}, \quad (15)$$

where

$$\langle \hat{I}(t'')\hat{I}(t')\rangle_S = \frac{1}{2}\langle \hat{I}(t'')\hat{I}(t')\rangle + \frac{1}{2}\langle \hat{I}(t')\hat{I}(t'')\rangle,$$

$$\langle \hat{I}(t'')\hat{I}(t')\rangle_A = \frac{1}{2}\langle \hat{I}(t'')\hat{I}(t')\rangle - \frac{1}{2}\langle \hat{I}(t')\hat{I}(t'')\rangle.$$

For an undamped oscillator the commutator is given by

$$\hat{x}(t')\hat{x}(t) - \hat{x}(t)\hat{x}(t') = \frac{\hbar}{2M\Omega}\exp[-i\Omega(t'-t)]$$
$$- \exp[i\Omega(t'-t)].$$

At the same time in the general case, the averaged commutator can be expressed by the system response, which we calculate using the classical equations of motion:

$$\langle \hat{x}(t')\hat{x}(t) - \hat{x}(t)\hat{x}(t')\rangle = i\hbar\alpha(t-t')$$
$$= \frac{\hbar}{2M\Omega^1}\exp[-i\Omega^1(t'-t)]$$
$$- \exp[i\Omega^1(t'-t)]\exp\left[-\frac{(t-t')\gamma}{2}\right], \quad t > t'.$$

In the same fashion, we determine the symmetrized correlator using the FDT

$$\langle \hat{x}(t')\hat{x}(t) + \hat{x}(t)\hat{x}(t')\rangle = 2\int \frac{d\omega}{2\pi}\exp[-i(t-t')]$$
$$\times 2\hbar\alpha''(\omega)\left\{\frac{1}{2} + \frac{1}{\exp(\hbar\omega/T_x - 1)}\right\},$$

where

$$\alpha''(\omega) = \frac{\gamma}{4M\Omega^1}\left\{\frac{1}{(\omega-\Omega^1)^2 + \gamma^2/4} - \frac{1}{(\omega+\Omega^1)^2 + \gamma^2/4}\right\}$$
$$= \gamma\frac{\omega/M}{(\omega^2+\Omega^2)^2 - 4\omega^2\Omega^2}.$$

Under the condition $\tilde{\Omega} \ll \gamma$ as well as $\tau_{el} \gg \gamma^{-1}, T_x^{-1}$, we have

$$\langle x(0)x(t)\rangle = \lambda^2\left(\frac{2}{\gamma}\right)^4\left\{\langle \hat{I}(0)\hat{I}(t)\rangle_S - i\langle \hat{I}(0)\hat{I}(t)\rangle_A\frac{8T_x}{\hbar\gamma}\right\}. \quad (16)$$

It follows that the current correlator measured with an ammeter is made up of both the symmetrized current correlator (whose coefficient can be derived from the classical equation of the circuit motion), and the antisymmetrized correlator $\langle \hat{I}(0)\hat{I}(t)\rangle_A$, whose share depends, in particular, on the temperature $T_x$.

Let us write out explicit expressions for the current correlators, assuming the conductance to have a single first order pole (in the case of several poles, we should sum the corresponding terms).

The antisymmetrized correlator coincides with the differential conductance to the accuracy of the derivative and coefficient:

$$\langle \hat{I}(0)\hat{I}(t)\rangle_A = iG\int \frac{d\omega}{2\pi}\sin(\omega t)\frac{\hbar\omega\omega_0^2}{\omega^2+\omega_0^2}$$
$$= \frac{i}{2}G\hbar\omega_0^2\exp(-\omega_0 t). \quad (17)$$

At equilibrium the expression for the symmetrized correlator

$$\langle \hat{I}(0)\hat{I}(t)\rangle_S = G\int \frac{d\omega}{\pi}\cos(\omega t)\frac{\hbar\omega\omega_0^2}{\omega^2+\omega_0^2}\frac{1}{\exp(\hbar\omega/2T_e)-1} \quad (18)$$

can be reduced to analytical expressions in different limiting cases. At $2\pi Tt/\hbar \gg 1$

$$\langle \hat{I}(0)\hat{I}(t)\rangle_S = G\hbar\omega_0^2\left\{\frac{\exp(-\omega_0 t)}{4\sin(\hbar\omega_0/2T)}\cos\frac{\hbar\omega_0}{2T}\right.$$
$$\left. - \frac{2\pi}{(\hbar\omega_0/T)^2 - (2\pi)^2}\exp\left(-\frac{2\pi Tt}{\hbar}\right)\right\}.$$

At $2\pi T \gg \hbar\omega_0$

$$\langle \hat{I}(0)\hat{I}(t)\rangle_S = G\hbar\omega_0^2\left\{\exp(-\omega_0 t)\frac{T}{2\hbar\omega_0}\right.$$
$$\left. + \frac{1}{2\pi}\ln\frac{1}{1-\exp(-2\pi Tt/\hbar)}\right\}.$$

At $\omega_0 t \gg 1$, $2\pi T \ll \hbar\omega_0$

$$\langle \hat{I}(0)\hat{I}(t)\rangle_S = -G\frac{\pi T^2}{2\hbar\sinh^2(\pi Tt/\hbar)}.$$

At $\omega_0 t \gg 1$, $2\pi T \ll \hbar\omega_0$, $Tt \ll \hbar$

$$\langle \hat{I}(0)\hat{I}(t)\rangle_S = -G\frac{\hbar}{2\pi t^2}.$$

It is this case that is the most typical for the appearance of vacuum fluctuations; in this limit the symmetrized correlator decreasing as a power can easily be separated from the antisymmetrized correlator decreasing as an exponent. It is also clear that the smaller the ratio $8T_x/\hbar\gamma$ characterizing the ammeter, the better the separation. If the ratio is rather small, the contribution of the symmetrized current correlator is always dominant.

The problem considered in this section is probably similar to that of the phase breaking of electrons by an electric field. This phase breaking does occur at zero temperature, but for basic reasons, it can hardly be calculated by the substitution of the symmetrized field correlator into the expressions for the phase breaking in a classical fluctuating field. This assertion can be proved by analysing diagrams describing the damping of the 'Cooperon', where a certain class of the diagrams, which is substantial for the phase breaking is zero in the quantum limit at zero temperature due to specific cancellations.

The author is thankful to M Feĭgel'man, S V Iordanskiĭ, V Kravtsov, L Levitov, V Fal'ko, Yu Belousov, and G Blatter for stimulating discussions.

This work was supported by the Russian Foundation for Basic Research (grant No. 96-02-19568) and the Swiss National Foundation.